\documentclass[a4paper,11pt]{elsarticle}
\usepackage[utf8]{inputenc}
\usepackage[margin=1in]{geometry}
\usepackage{amsmath}
\usepackage{amssymb}
\usepackage{graphicx}
\graphicspath{ {./Plots/} }
\usepackage{xcolor}
\usepackage{hhline}
\usepackage{adjustbox}
\usepackage{subcaption}

\title{Percolation of Domain Walls in the Two-Higgs Doublet Model}
\author{Richard A. Battye, Steven J. Cotterill, Eva Sabater Andres and Adam K. Thomasson}
\address{Jodrell Bank Centre for Astrophysics, Department of Physics and Astronomy, University of Manchester, Oxford Road, Manchester M13 9PL}

\begin{document}

\begin{abstract}
Domain walls formed during a phase transition in a simple field theory model with $\mathbb{Z}_2$ symmetry in a periodic box have been demonstrated to annihilate as fast as causality allows and their area density scales $\propto t^{-1}$. We have performed numerical simulations of the dynamics of domain walls in the Two-Higgs Doublet Model (2HDM) where the potential has $\mathbb{Z}_2$ symmetry in two spatial dimensions. We observed significant differences with the standard case. Although the extreme long-time limit is the same for the $\approx 10^{5}$ sets of random initial configurations analysed, the percolation process is much slower due to the formation of long-lived loops. We suggest that this is due to the build up of superconducting currents on the walls which could lead ultimately to stationary configurations known as Kinky Vortons. We discuss the relevance of these findings for the production of Vortons in three spatial dimensions.
\end{abstract}

\maketitle

\section{Introduction}\label{sec:Intro}

Topological defects are expected to form at cosmological phase transitions via the Kibble mechanism whenever the homotopy groups of the vacuum manifold are non-trivial. They can lead to many interesting cosmological consequences, see for example ref.~\cite{Vilenkin278400}.

The focus of this {\it letter} is on domain walls which can form when the vacuum manifold is disconnected, that is, when $\pi_0({\cal M})\ne I$ where ${\cal M}$ is the vacuum manifold. The simplest model that admits domain walls is a $\mathbb{Z}_2$ symmetric model with Lagrangian density
\begin{equation}
\label{eq:Lagrangian_Goldstone}
\mathcal{L} = \frac{1}{2} \partial_\mu \phi\partial^{\mu} \phi - \frac{\lambda}{4}  (\phi^2 - \eta^2)^2\,,
\end{equation}
where $\eta$ and $\lambda$ are positive constants and $\phi$ is a real scalar field. By a choice of length and energy units we can scale $\eta$ and $\lambda$ out of the problem and, hence, without loss of generality we can set $\eta=\lambda=1$. Large-scale numerical simulations of this model in periodic boxes in three spatial dimensions, for example ref.~\cite{DW_scaling}, have demonstrated that the area density of walls scales $\propto t^{-1}$ where $t$ is the simulation time, implying a scaling length $L\propto t$; the density of walls $\propto \sigma A/V$ where $A\propto L^2$ is the area and $V\propto L^3$. This process has been modelled semi-analytically~\cite{Avelino:2005kn,Leite:2012vn,Martins:2016ois} and has been demonstrated in a number of other models \cite{Oliveira:2004he,Battye:2006pf,Leite:2011sc}. 

An alternative approach to studying the dynamics of domain walls is to observe the percolation to the true vacuum in the long-time limit. This has been investigated in the context of condensed matter systems, that is, those with non-relativistic dynamics. For the Ising model in two spatial dimensions it has been predicted analytically~\cite{prob} that a fraction of $\approx 0.64$ of random initial configurations in a periodic box would be expected to reach the true vacuum and the rest will reside in various striped states which wind around the toroidal geometry induced by the periodic boundaries. Due to the general nature of the calculation, which is not particularly sensitive to the details of the fields nor the time evolution, it is interesting to understand the robustness, or otherwise, of this prediction.

One can add a complex scalar field $\sigma$ to the simple $\mathbb{Z}_2$ model to yield a $\mathbb{Z}_2 \times U(1)$ symmetric Lagrangian,
\begin{equation}
\mathcal{L} = \partial_\mu \phi\partial^\mu \phi + \partial_\mu \sigma\partial^\mu \bar\sigma - \frac{\lambda_\phi}{4} \left (\phi^2 - \eta^2_\phi \right )^2 - \frac{\lambda_\sigma}{4} \left (|\sigma|^2 - \eta^2_\sigma \right )^2  - \beta \phi^2 |\sigma|^2\,,
\label{eq:KV_lagrangian}
\end{equation}
where, by scaling of units as before, we can set $\eta_\phi=\lambda_\phi=1$.
This model admits superconducting wall solutions and these can be calculated analytically~\cite{Hodges:1988qg,Battye_2008_KV} when $2\beta = \lambda_\phi = \lambda_\sigma$, which fixes all the parameters except $\eta_\sigma$ up to the rescaling of length and energy.

It has been shown that one can construct stable circular superconducting loops of domain wall known as Kinky Vortons ~\cite{Battye_2008_KV,Battye_FEKV}, which are two-dimensional analogues of Vortons. In this scenario domain walls emerge from the breaking of the $\mathbb{Z}_2$ symmetry and the $\sigma$-field acts as a superconducting condensate. Although these structures are two-dimensional they can provide relevant insight into the study of the three-dimensional case of Vortons. 

Here, we study the formation and evolution of domain walls in the Two-Higgs Doublet model (2HDM) which is a well studied extension of the Standard Model (SM)~\cite{Branco:2011iw}. As we will explain in the proceeding sections it is natural for such a model to have an extra $\mathbb{Z}_2$ symmetry in addition to those of the SM. The breaking of this extra symmetry can lead to the formation of domain walls~\cite{Battye_2011_vacTop,Eto:2018hhg,Battye:2020jeu,Battye_2021_Dom_sims,Law:2021ing} and the evolution of these domain walls has been seen~\cite{Battye_2021_Dom_sims} to violate the standard scaling law found for the simple $\mathbb{Z}_2$ model. It was suggested that this could be due to the existence of superconducting currents on the walls which slow down the collapse of loops and could, in the end, lead to the formation of structures similar to Kinky Vortons. We will do this by performing numerical simulations of the 2HDM similar to those in ref.~\cite{Battye_2021_Dom_sims}, except we will investigate what happens beyond the Light Crossing Time (LCT) - the time taken for signals travelling at the speed of light to propagate across half the size of the periodic box - with a view to calculating the long-time field configuration and to investigate the eventual percolation to the true vacuum, or otherwise, in keeping with the spirit of ref.~\cite{prob}. In order to compare, we will also provide similar results for the simple $\mathbb{Z}_2$ and $\mathbb{Z}_2 \times U(1)$ models. 

\section{2HDM with $\mathbb{Z}_2$ Symmetry}\label{sec:2HDM_theory}

The general Lagrangian of the 2HDM can be written as
\begin{equation}
\mathcal{L} = (\partial^\mu \Phi_1)^\dagger (\partial_\mu \Phi_1) + (\partial^\mu \Phi_2)^\dagger (\partial_\mu \Phi_2) - V(\Phi_1, \Phi_2),
\label{eq:2HDM_lag}
\end{equation}
where $\Phi_{1,2}$ are two complex doublets. Restricting the potential to be symmetric under the $\mathbb{Z}_2$ transformation, $\Phi_1 \rightarrow \Phi_1, \Phi_2 \rightarrow - \Phi_2$, it takes the form
\begin{align}
	V(\Phi_1, \Phi_2) = & -\mu_{1}^2 (\Phi_1^\dag \Phi_1) - \mu_2^2 (\Phi_2^\dag \Phi_2) \nonumber + \lambda_1 (\Phi_1^\dag \Phi_1)^2 + \lambda_2 (\Phi_2^\dag \Phi_2)^2 \nonumber + \lambda_3 	(\Phi_1^\dag \Phi_1)(\Phi_2^\dag \Phi_2) \nonumber \\
	& + (\lambda_4 - |\lambda_5|) \left [\mathrm{Re}(\Phi_1^\dag \Phi_2)  \right ]^2 + (\lambda_4 + |	\lambda_5|) \left [\mathrm{Im}(\Phi_1^\dag \Phi_2)  \right ]^2\,.
\label{eq:2HDM_Potential}
\end{align}
The $\mathbb{Z}_2$ symmetric case is a sub-class of the more general 2HDM which is often assumed in phenomenological studies due to its ability to prevent tree-level flavour changing neutral currents (FCNCs), see, for example, ref.~\cite{Branco:2011iw}.

 Alternatively, one can express the potential of the model in a \textit{bi-linear field-space formalism }~\cite{Battye_2011_vacTop, Ivanov:2006yq, Ivanov_2008, Maniatis_2006} where one term contains the quadratic mass parameters and the second term contains the quartic couplings, 
\begin{equation}
    V = -\frac{1}{2}M_\mu R^\mu + \frac{1}{4}L_{\mu \nu} R^\mu R^\nu.
\end{equation}
The bi-linear field vector, $R^\mu$, has the form 
\begin{equation}
    R^\mu = \Phi^\dag \sigma^\mu \Phi = \left(\begin{matrix}
        \Phi_1^\dag \Phi_1 + \Phi_2^\dag \Phi_2\\
        \Phi_1^\dag \Phi_2 + \Phi_2^\dag \Phi_1\\
        -i \left[\Phi_1^\dag \Phi_2 - \Phi_2^\dag \Phi_1\right]\\
        \Phi_1^\dag \Phi_1 - \Phi_2^\dag \Phi_2
    \end{matrix}\right),
\end{equation}
and is invariant under an electroweak transformation. For the specific forms of $M_\mu$ and $L_{\mu\nu}$ see, for example, ref.~\cite{Battye_2011_vacTop}. By the introduction of the $SU(2)_L$ invariant object, $\Phi_1^T i \sigma^2 \Phi_2$, the vector $R^\mu$ can be promoted to a null 6-vector where~\cite{Battye_2011_vacTop}
\begin{equation}
    R^4 = \Phi_1^T i \sigma^2 \Phi_2 - \Phi_2^\dagger i \sigma^2 \Phi_1^*, \qquad R^5 = -i\left(\Phi_1^T i \sigma^2 \Phi_2 + \Phi_2^\dagger i \sigma^2 \Phi_1^*\right),
\end{equation}
which allows for the study of further topological structure in the fields. We may also note that $R^\mu R_\mu = (R^4)^2 + (R^5)^2$.

The $\mu_i$ and $\lambda_i$ parameters of the potential of (\ref{eq:2HDM_Potential}) can be expressed in terms of the masses of the five physical scalar particles, $M_h, M_H, M_A \text{ and } M_{H^\pm}$, the vacuum expectation value of the SM, $v_{SM}$, the mixing angle between the CP-even fields of the model, $\alpha$, and the ratio of the vacuum expectation values, $\tan\beta=v_2/v_1$, of the two fields. See the appendix of ref.~\cite{Battye:2020jeu} for formulae connecting the two different formulations. In the simulations presented here, we limit ourselves overall to $M_H = M_A = M_{H^\pm} = 200~\text{GeV}$ and $\tan\alpha=\tan\beta=0.85$, with $M_h = 125~\text{GeV}$ and $v_{\rm SM} = 246~\text{GeV}$ being fixed by experiment~\cite{Workman:2022ynf}. We also rescale the potential for dimensionless length and energy, using the fixed values of $M_h$ and $v_{\rm SM}$.

Contrary to the simple $\mathbb{Z}_2$ model, the domain walls do not appear in the fields but in the components of $R^\mu$. In particular, solving of the field equations in one-dimension and performing full dynamical simulations~\cite{Battye_2011_vacTop,Battye_2021_Dom_sims,sassi2023domainwallstwohiggsdoubletmodel}, one finds that $R^1$ exhibits a plus to minus transition across the centre of the wall with condensate-like structures in $R^2$ and local maxima in $R^3$ on the walls. In addition, winding of these condensate structures have also been observed in the extended components of $R^4$ and $R^5$. In these respects it has similarities with the superconducting wall solutions in the $\mathbb{Z}_2 \times U(1)$ model.

\section{Numerical Techniques}\label{sec:Num_tech}

Simulations were performed in two spatial dimensions on  regular grids of $P^2$ points with dimensionless grid spacing $\Delta x = 0.5$. Due to the number of simulations required for the fractional analysis presented here $(\approx 10^5)$, small grids with $P=256$ or $P=512$ were used. The equations of motion were discretized using central finite difference methods, approximating temporal derivatives to 2nd order and spatial derivatives to 4th order. The fields were initialised with random initial conditions (RIC) for each grid point, chosen to approximate the outcome of a 2nd-order phase transition, and then subsequently evolved in time using dimensionless timestep $\Delta t = 0.1$.

The use of RIC creates unphysically large initial gradients between adjacent points. Thus, to better model a phase transition, an initial short period of damped evolution was employed to smooth the large gradients. During the damping period, $\epsilon$ - the coefficient of the field time derivative term - was set to unity and subsequently set to zero once the damping period was complete: after 300 timesteps had passed. Periodic boundary conditions were used which imposes a LCT given by $P\Delta x/(2\Delta t)$, after which a signal could have interacted with itself and, therefore, cannot be considered to be representative of the physical dynamics in terms of modelling the scaling behaviour. Nevertheless, the simulations presented here run beyond this limit in order to mimic the scenario discussed in ref.~\cite{prob}.

An important quantity to track during the simulations is the number of domain walls, $N_{\rm DW}$, which has been shown to act as a reliable scaling proxy for the area density \cite{PhysRevD.58.103501}. This was calculated by, at each timestep of the evolution, iterating through the simulation space and incrementing the number of domain walls whenever the sign of the field configuration flipped between adjacent grid points: $\phi$ in the case of the $\mathbb{Z}_2$ and $\mathbb{Z}_2 \times U(1)$ models; $R^1$ in the case of the 2HDM. Adjacent points were only compared in the positive direction of each axis so as not to double count. 

\begin{figure}
    \centering
        \includegraphics[width=0.185\textwidth]{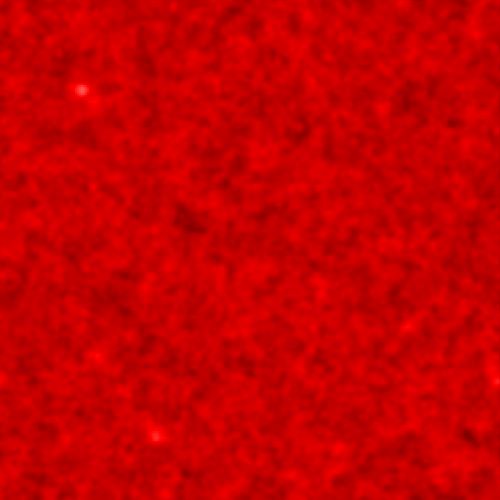}\hfill
        \includegraphics[width=0.185\textwidth]{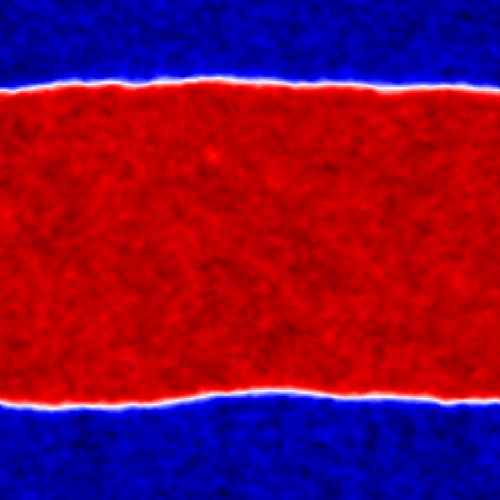}\hfill
        \includegraphics[width=0.185\textwidth]{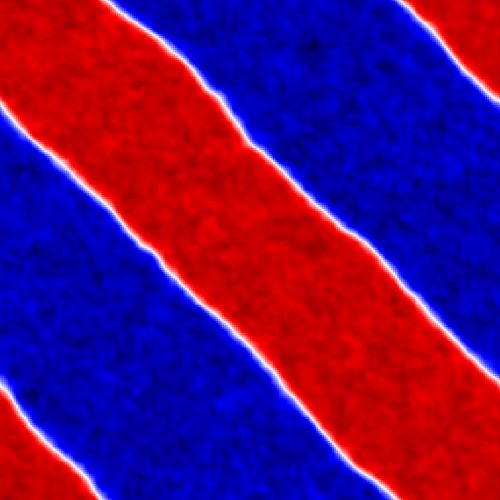}\hfill
        \includegraphics[width=0.185\textwidth]{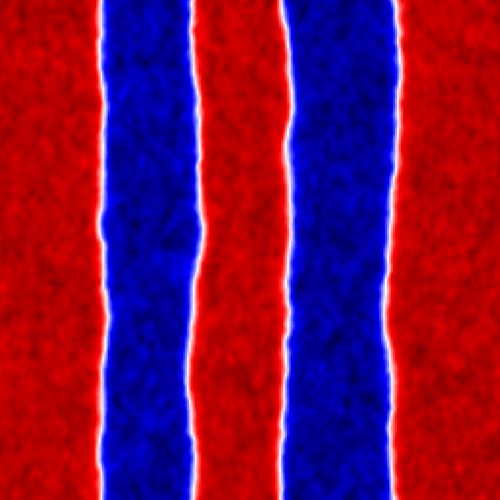}\hfill
        \includegraphics[width=0.185\textwidth]{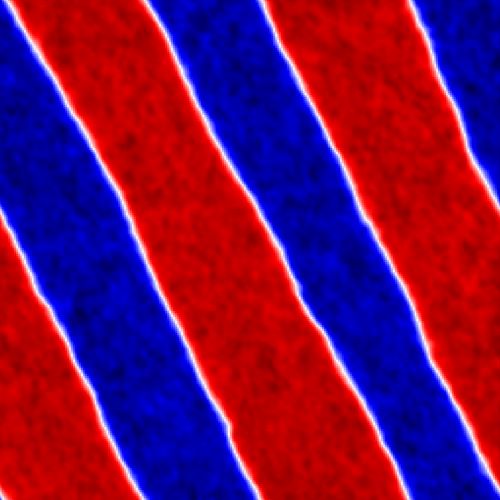}\hfill
        \includegraphics[height=0.185\textwidth, width=0.0175\textwidth]{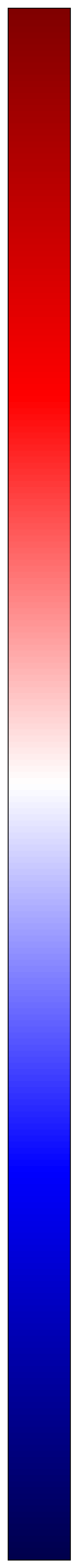}
        \includegraphics[height=0.185\textwidth]{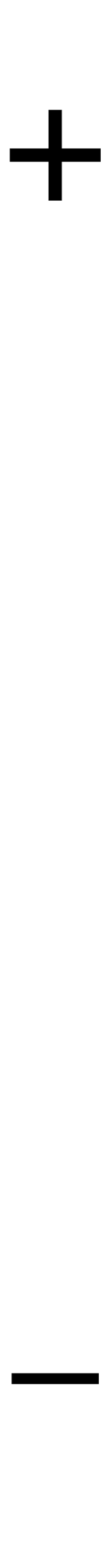}
\caption{Final states of $R^1$ seen in the simulations of the 2HDM. A colour mapping is used where red represents the positive vacuum expectation value, blue the negative and white the interpolating walls. From left to right: single domain (SD), single stripe, diagonal stripe, double stripe and double diagonal stripe.}
\label{fig:states}
\end{figure}

The toroidal geometry imposed by the periodic boundary conditions allows different types of final state and these are illustrated for the case of the 2HDM in Fig.~\ref{fig:states}; similar patterns were found in the cases of  $\mathbb{Z}_2$ and $\mathbb{Z}_2 \times U(1)$ models. These states include a single vacuum and two types of striped states which wind differently around the torus. For each type of winding there can be integer numbers of twists and we only show up to winding number two, given that greater winding was not observed in our simulations. 

Our objective is to make estimates of the fraction of simulations from an ensemble which lead to a box containing a single domain, $f_{\rm SD}$, and the fractions of different striped states. In order to do this we have used two methods. The first, which we will denote method A, uses the number of grid points which exist in the positive or negative vacua to estimate the proportion of each domain value present as a function of time. Once the minority vacua's proportion decreases below a predefined fraction (determined to be 0.05 for best classification) the simulation is deemed to have reached a single domain state. Alternatively, method B uses histograms of $N_{\rm DW}$ during simulations to delineate different kinds of configurations - realizations with $N_{\rm DW}\approx 0$ will be assumed to be in a single vacuum state whereas the various striped states will correspond to specific values of $N_{\rm DW}$ governed by the size of the box.

\section{Simulation Results}\label{sec:Results}

\begin{figure}
    \centering
    \includegraphics[width=0.75\linewidth]{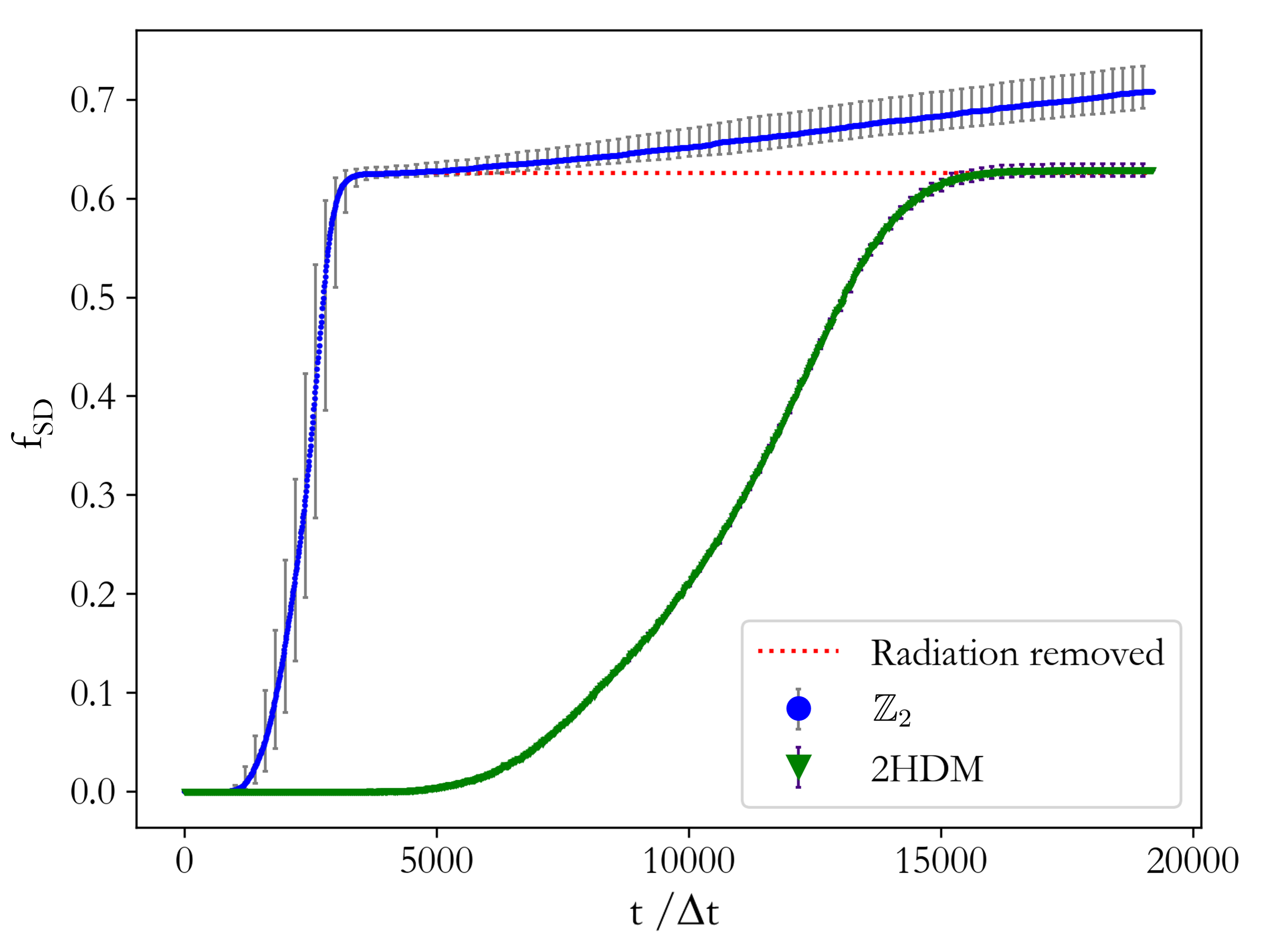}
    \caption{Fraction of realizations which form a single domain, $f_{\rm SD}$, for a sample of $10^4$ for the simple $\mathbb{Z}_2$ model (blue) and the 2HDM with $\mathbb{Z}_2$ symmetry (green) as a function of the number of timesteps. In addition we have included the case of the simple $\mathbb{Z}_2$ where radiation is removed (red).  Simulations were performed on grids with $P=512$ and were terminated at 15 LCTs. It is clear that the asymptotic values are similar for the green and red curves, but that the relaxation time is much longer.}
    \label{fig:Gold_2HDM}
\end{figure}

In Fig.~\ref{fig:Gold_2HDM} we present the calculated value $f_{\rm SD}$ as a function of time using method A for the 2HDM with $\mathbb{Z}_2$ symmetry and $P=512$, compared to that for the simple $\mathbb{Z}_2$ model. This has been done for $10^{4}$ different sets of RIC in each case, run to 15 LCTs, and the error bars denote the Poisson errors. The value asymptotes to $f_{\rm SD}\approx 0.62$ in the long-time limit for the case of the 2HDM, which is close to the value predicted in ref~\cite{prob}. In the case of the simple $\mathbb{Z}_2$ model - and indeed also in the $\mathbb{Z}_2\times U(1)$ model\footnote{We note that charge and current do appear to become localised on the walls. These appear to have little impact on the dynamics, which were similar to the simple $\mathbb{Z}_2$ model for the parameters that we used and are in stark contrast to what we report here in the 2HDM. We shall touch upon the similarity of the two simpler models later in our discussion.}, which is not presented - $f_{\rm SD}$ increases quickly to a value close to that of the 2HDM before gradually increasing. This appears to be related to pockets of radiation  which perturb walls that are trying to settle down into a striped state. If the perturbations are sufficiently large one can imagine they could destabilise the configuration enough to allow it to transition to a single vacuum state. This does not appear to be the case in the 2HDM. We suggest that this might be due to the existence of massless radiation, associated with the Goldstone bosons that arise when the global SM symmetries are broken. In the cases of the two simpler models, there is only massive radiation in the vacuum, related to perturbations in $\phi$ and $|\sigma|$, with no Goldstone bosons as the $U(1)$ symmetry is unbroken outside the walls. If we remove the radiation in the simple $\mathbb{Z}_2$ simulations, by periodically setting $\dot\phi = 0$ for a short number of timesteps, then it does settle down to a very similar value of $f_{\rm SD}$. It is clear that the relaxation time to the asymptotic value of $f_{\rm SD}$ is very much longer in the case of the 2HDM which we will return to later in the discussion.

\begin{figure}
    \begin{minipage}{0.6\textwidth}
        \centering
        \includegraphics[width=\columnwidth]{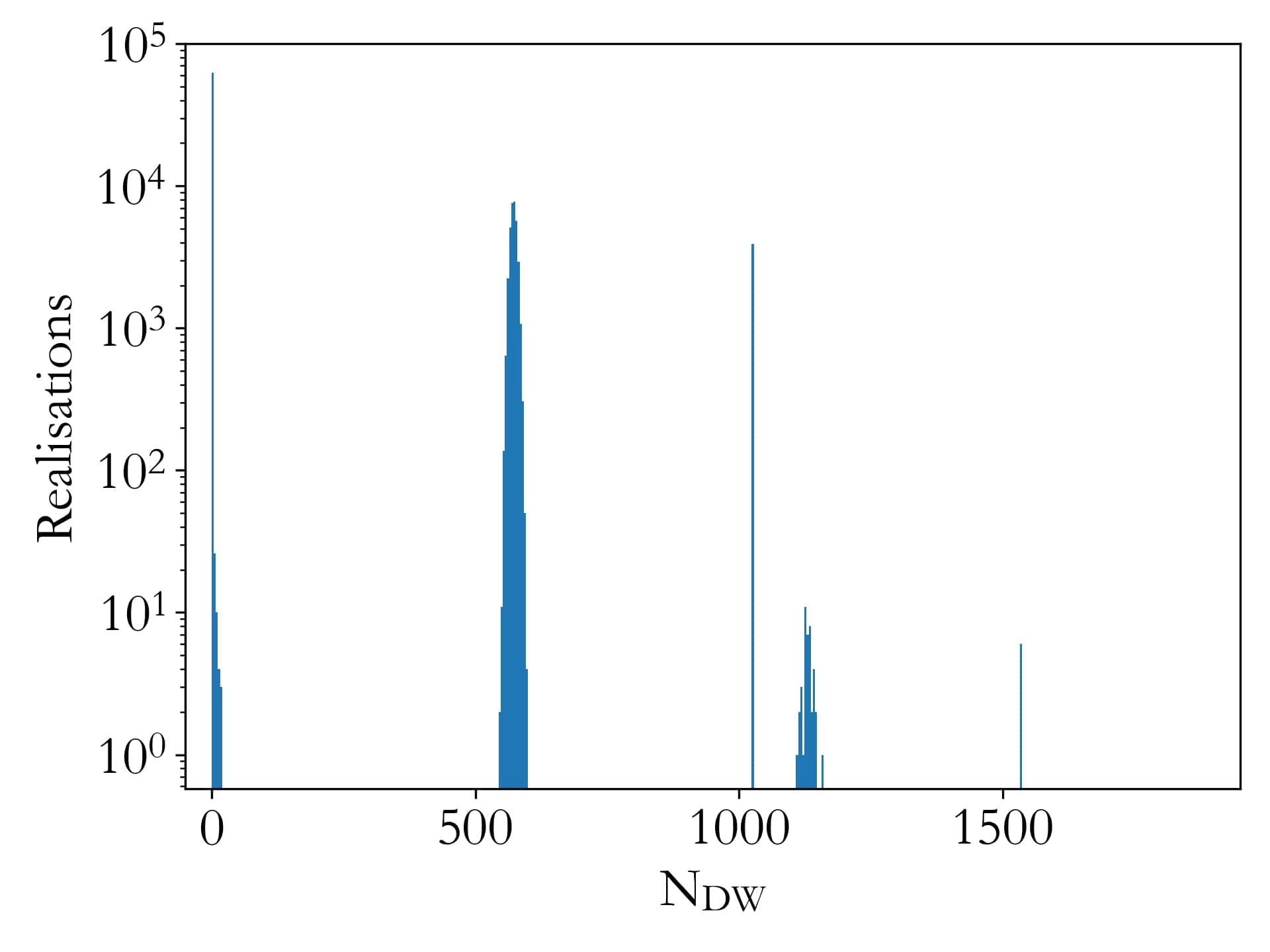}
    \end{minipage}
    \begin{minipage}{0.3\textwidth}
    \centering
    \small
    \begin{tabular}{| c | c |}
		\hline
		Final State &  Simulation \\
                        & Fraction \\
		\hhline{|=|=|}
		Single Domain &  0.62473\\
		Stripe &  0.33579\\
		Diagonal Stripe &  0.03900\\
		Double Stripe & 0.00042\\
		  Double Diagonal Stripe & 0.00006\\
		\hline
	\end{tabular}
    \end{minipage}
    \caption{Histogram of $N_{\rm DW}$ for $10^5$ realizations of the 2HDM with grids of $P=256$ (left). In each case the evolution has effectively ceased and simulations are in a stable final state. From left to right the five peaks of the histogram represent the five stable final states depicted in Fig.~\ref{fig:states} respectively. Each peak's fraction within the sample is given in the table (right).}
    \label{fig:2HDM_final_hist}
\end{figure}

It seems that method A gives a value of $f_{\rm SD}$ which is compatible with the analytic prediction in ref.~\cite{prob}, but that there could be issues associated with radiation, although this seems to be less of an effect in the case of the 2HDM. We now turn our attention to method B. In Fig.~\ref{fig:2HDM_final_hist} we present a histogram of $N_{\rm DW}$ calculated after 12 LCTs for $10^5$ realizations of RIC with $P=256$. A smaller grid size was chosen to allow more realizations to be performed for the 2HDM (where 8 individual fields are being evolved, vastly increasing the computational resources required). The histogram exhibits five distinct peaks which we can attribute, by inspection, to the five end states illustrated in Fig.~\ref{fig:states}. In particular we can estimate $f_{\rm SD}=0.625\pm 0.003$ where the uncertainty has been calculated assuming Poisson errors. We are also able to calculate the fractions associated with the other states which have similar errors. We are not able to accurately determine the number of double stripe and double diagonal stripe states, but they are clearly non-zero in that we find realizations where they exist. We do not find any realizations where there is anything other than the five states in Fig.~\ref{fig:states} in the final configuration after 12 LCTs, although other states with winding numbers greater than two would presumably be formed very rarely if the sample of realizations were larger.

\begin{figure}
    \centering
         \begin{subfigure}{0.32\textwidth}
              \includegraphics[width=\columnwidth]{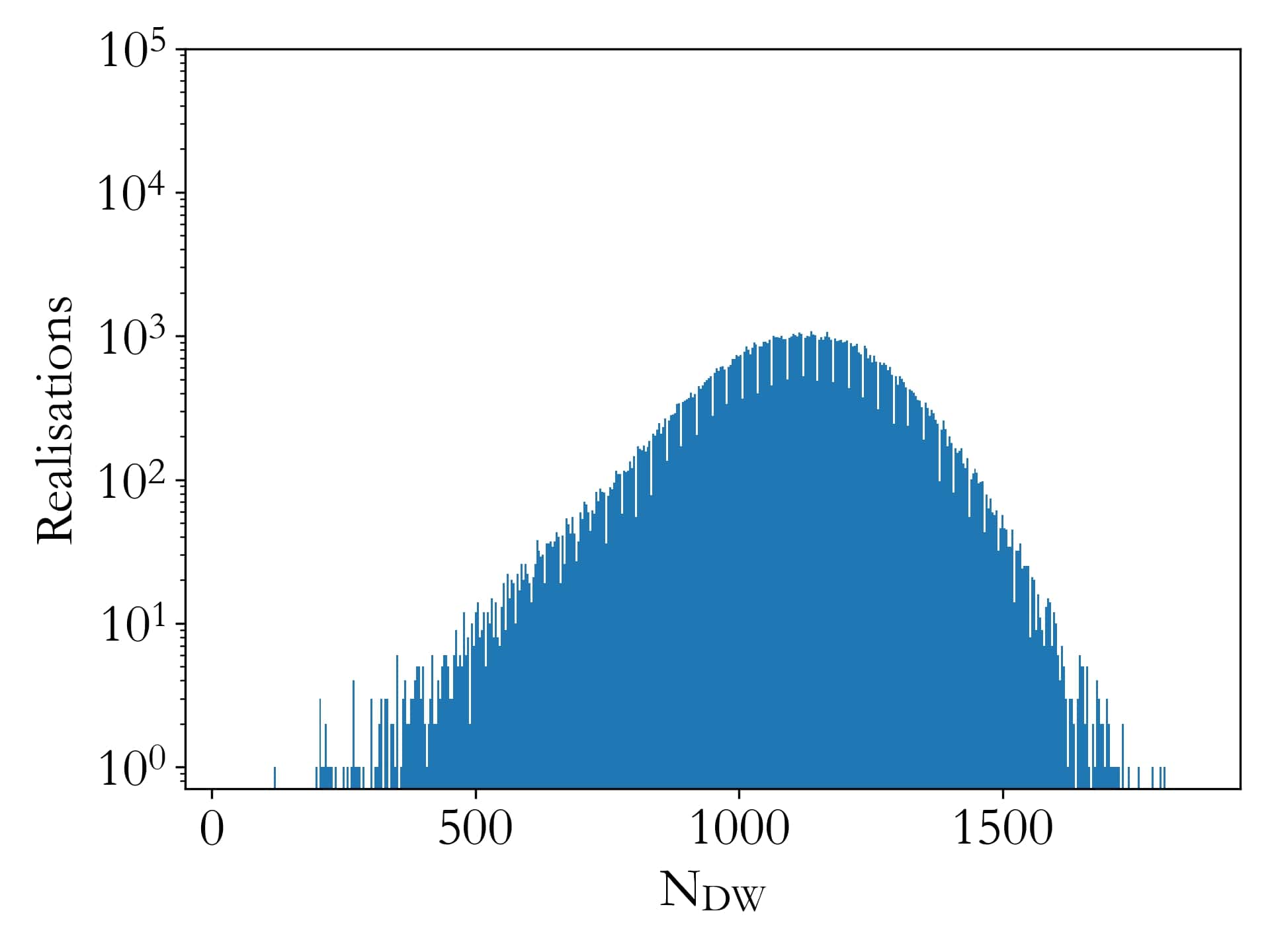}\\
              \includegraphics[width=\columnwidth]{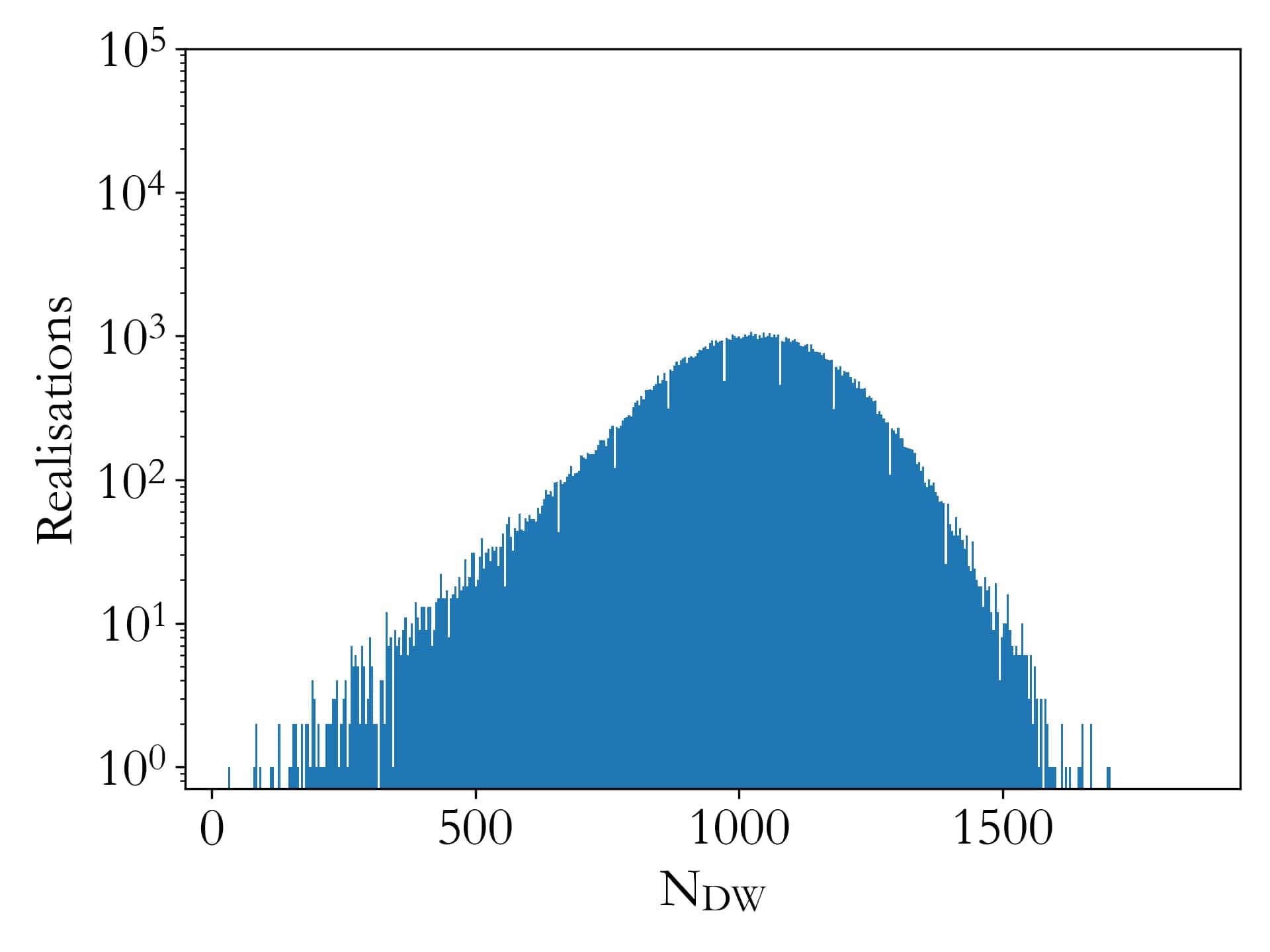}\\
              \includegraphics[width=\columnwidth]{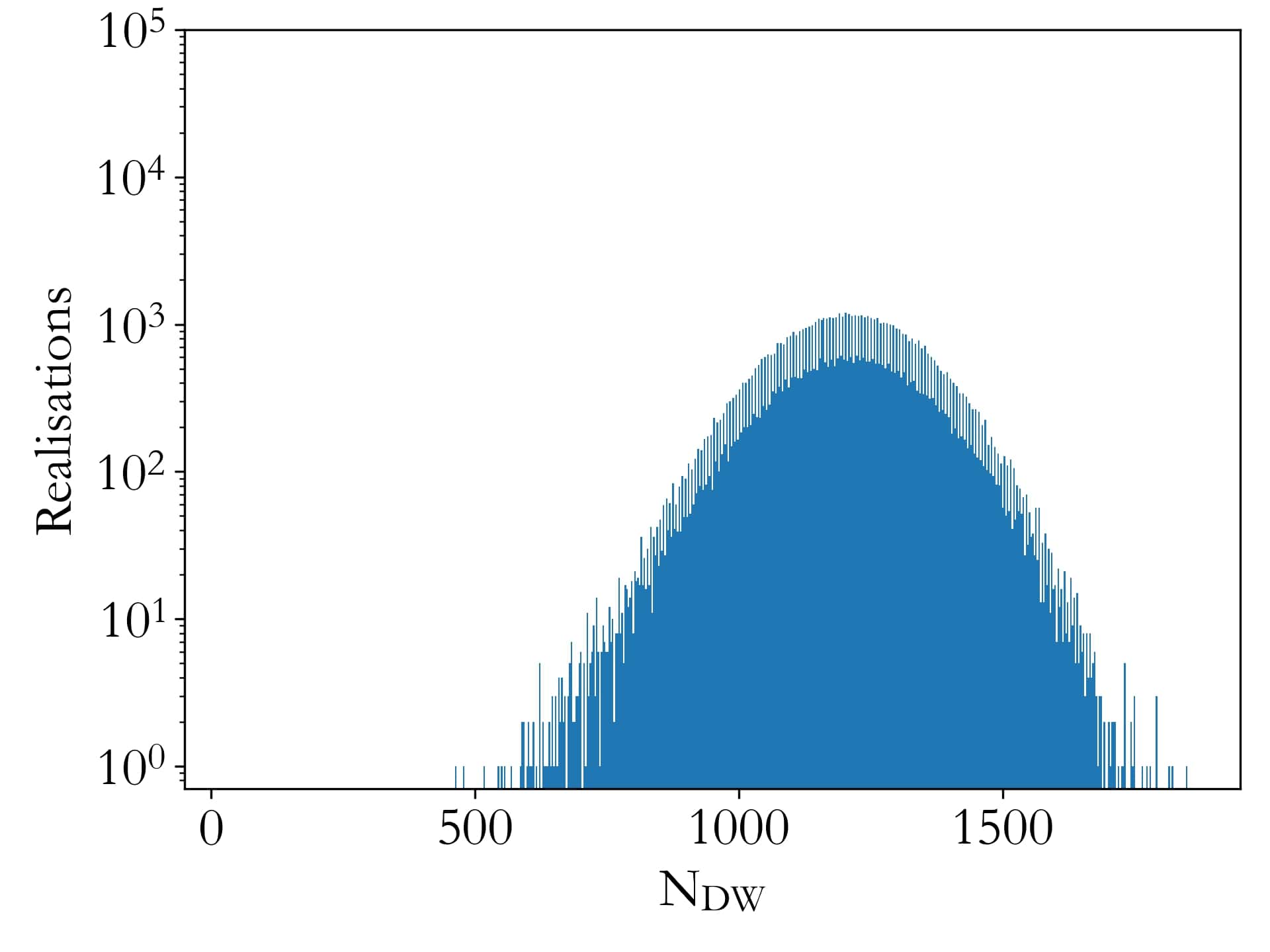}
         \caption{1 LCT}
         \label{fig:hists_1lct}
         \end{subfigure}
        \begin{subfigure}{0.32\textwidth}
              \includegraphics[width=\columnwidth]{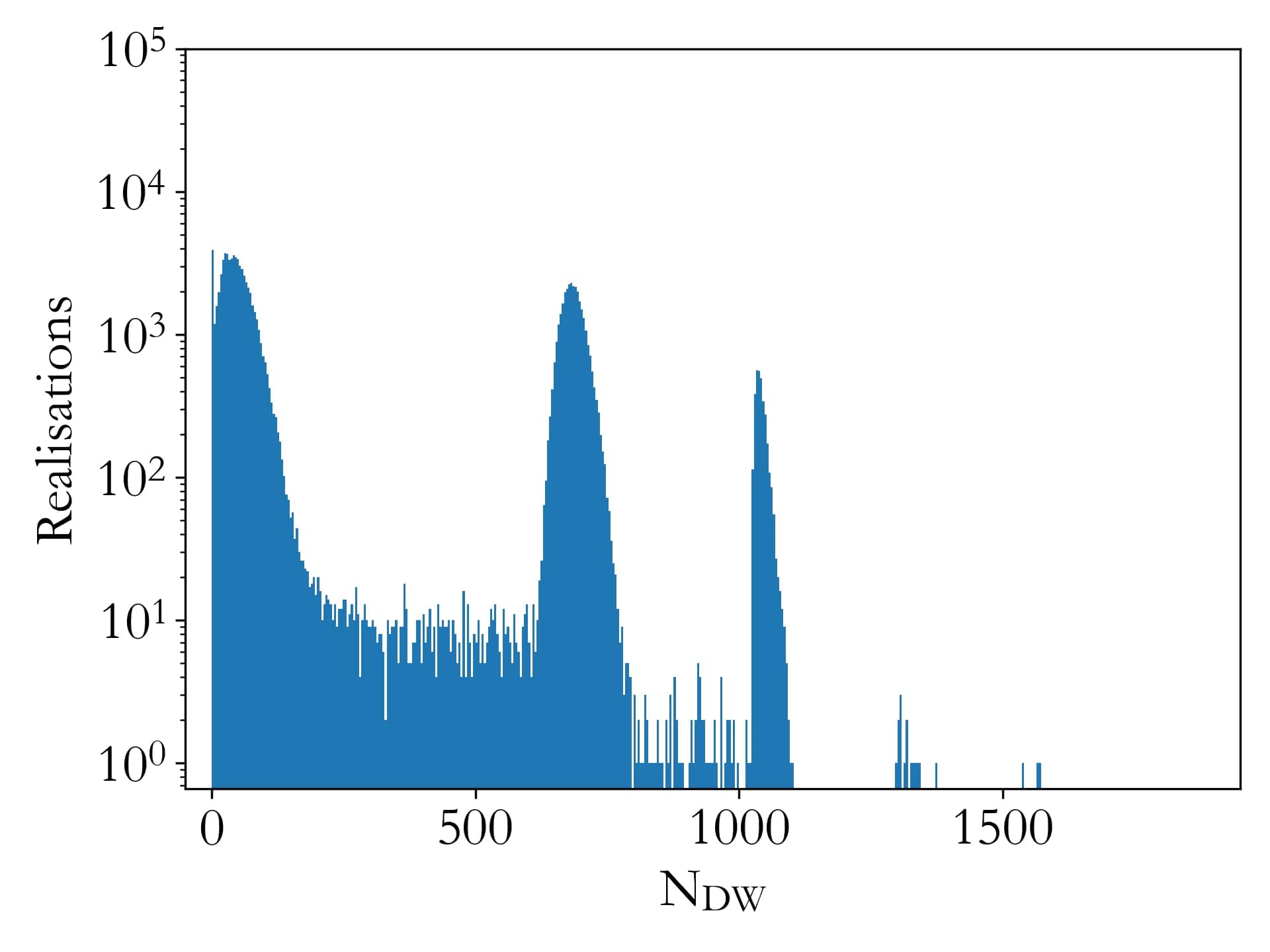}\\
              \includegraphics[width=\columnwidth]{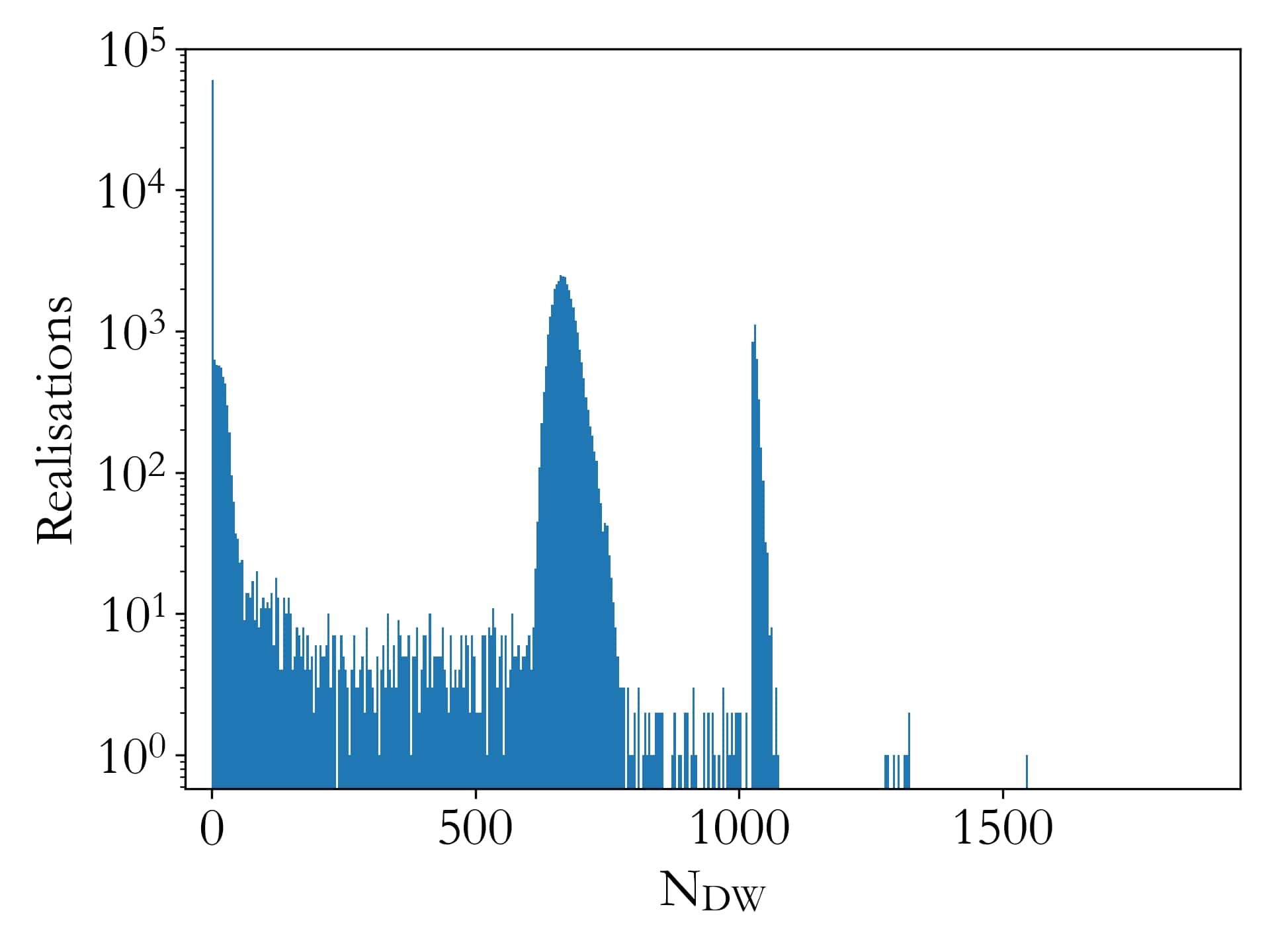}\\
              \includegraphics[width=\columnwidth]{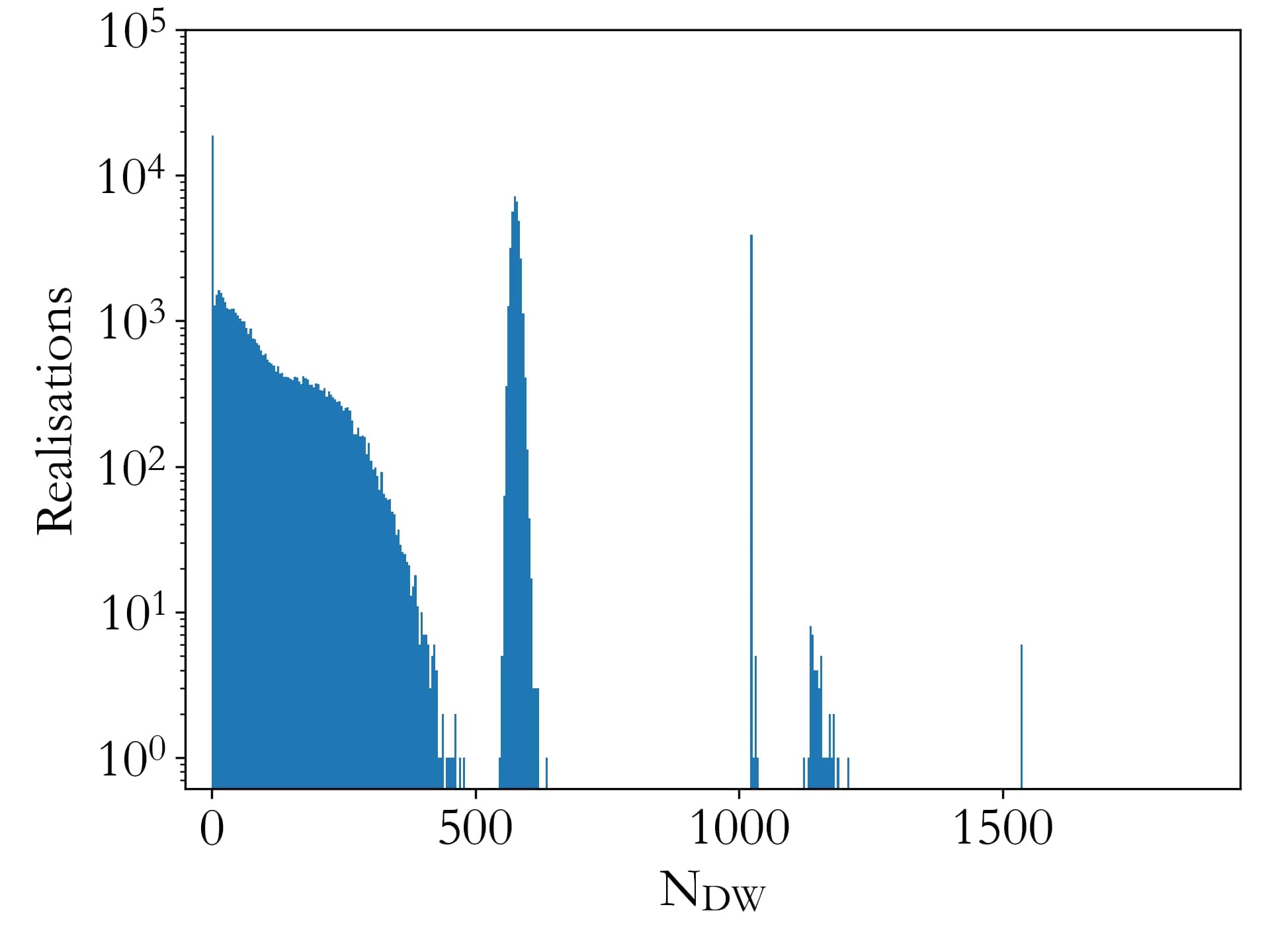}
         \caption{6 LCTs}
         \label{fig:hists_6lct}
         \end{subfigure}
        \begin{subfigure}{0.32\textwidth}
              \includegraphics[width=\columnwidth]{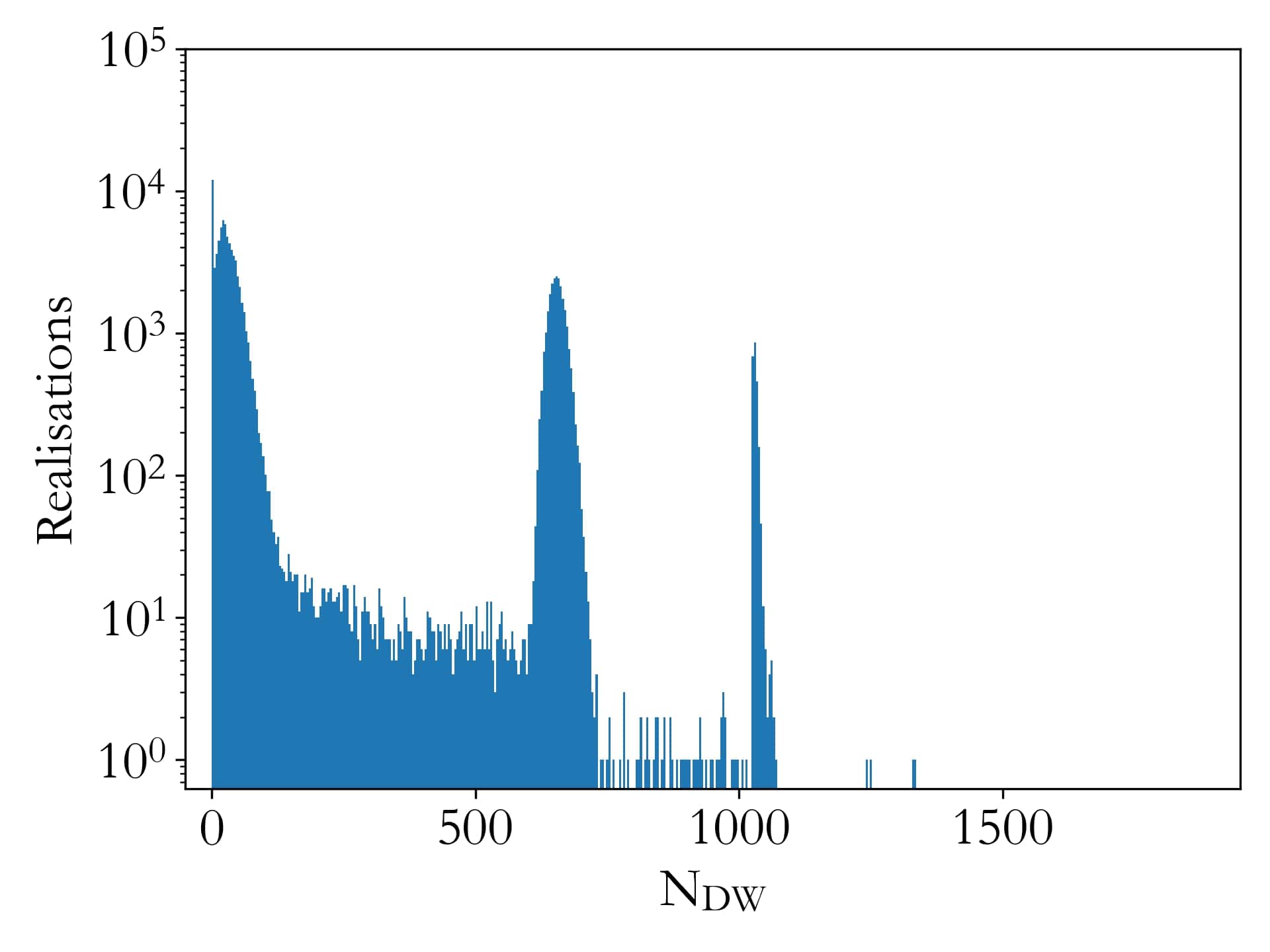}\\
              \includegraphics[width=\columnwidth]{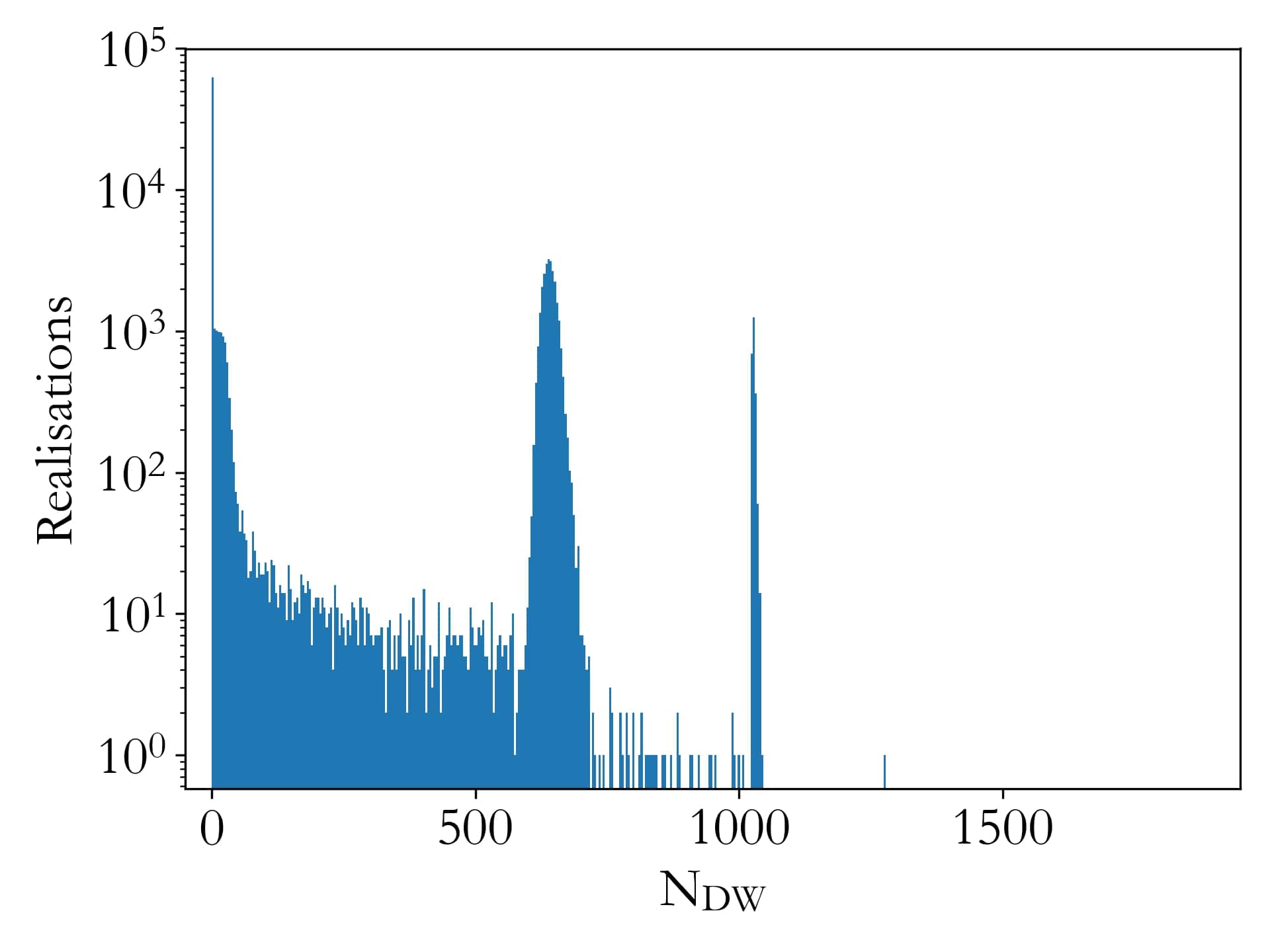}\\
              \includegraphics[width=\columnwidth]{hist_2HDM_12LCT_log.jpg}
         \caption{12 LCTs}
         \label{fig:hists_12lct}
         \end{subfigure}
    \caption{Histograms of $N_{\rm DW}$ (with bin width of 4) at three different times (1 LCT, 6 LCTs and 12 LCTs) during the evolution of a sample of $10^5$ simulations in grids of $P=256$. The rows are for three different models: the $\mathbb{Z}_2$ model (top), the $\mathbb{Z}_2 \times U(1)$ model (middle) and the 2HDM (bottom). The final histogram of the 2HDM is that of Fig.~\ref{fig:2HDM_final_hist}.}
    \label{fig:hists}
\end{figure}

Although the 2HDM appears to be ultimately in close agreement with the predictions of ref.~\cite{prob} in the long-time limit, the relaxation was seen to be much slower than in the simple $\mathbb{Z}_2$ model and indeed also the $\mathbb{Z}_2\times U(1)$ model. Fig. \ref{fig:hists} shows the equivalent of Fig.~\ref{fig:2HDM_final_hist} for the three models and at different times - 1 LCT, 6 LCTs and 12 LCTs within the sample of RIC. The distributions of $N_{\rm DW}$ are very similar after the first LCT for the cases of the simple $\mathbb{Z}_2$ and $\mathbb{Z}_2\times U(1)$ models, but even at this early stage one sees that there are significant differences with the 2HDM even though the initial conditions are very similar - this reflects the previously observed fact that the 2HDM simulations are not scaling at this point~\cite{Battye_2021_Dom_sims}. As the evolution continues the five distinct peaks emerge, corresponding to the stable final states.

There are two clear qualitative points that can be gleaned from examining Fig.~\ref{fig:hists} in the range $0\le N_{\rm DW}<800$:
\begin{itemize}
    \item In the cases of the simple $\mathbb{Z}_2$ and $\mathbb{Z}_2\times U(1)$ models the peaks are much less sharp, for example, characterized by configurations between the peak at $N_{\rm DW}\approx 0$ and the second peak. The final histogram in the case of the 2HDM is extremely sharp.
    \item The relaxation to the single vacuum is much slower in the case of the 2HDM. Modulo the effects of the radiation which is responsible for the lack of sharpness in the peaks, there are a very significant number of realizations in the case of 6 LCTs which have $N_{\rm DW}$ higher than the first peak.

\end{itemize}
This second observation, coupled with the results from Fig.~\ref{fig:Gold_2HDM} could suggest that there are long-lived loops forming in the case of the 2HDM.

\begin{figure}
\centering
    \begin{subfigure}[t]{0.23\textwidth}
        \includegraphics[width=\columnwidth]{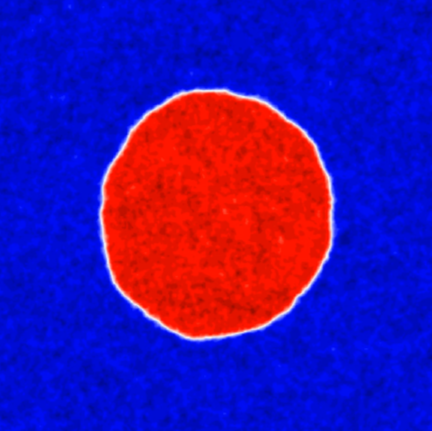}
        \caption{$R^1$}
    \end{subfigure}
    \begin{subfigure}[t]{0.23\textwidth}
        \includegraphics[width=\columnwidth]{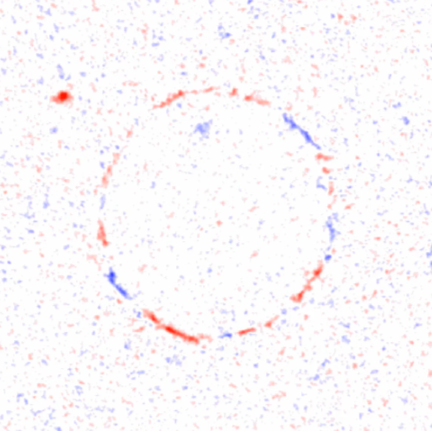}
        \caption{$R^2$}
    \end{subfigure}
    \begin{subfigure}[t]{0.23\textwidth}
        \includegraphics[width=\columnwidth]{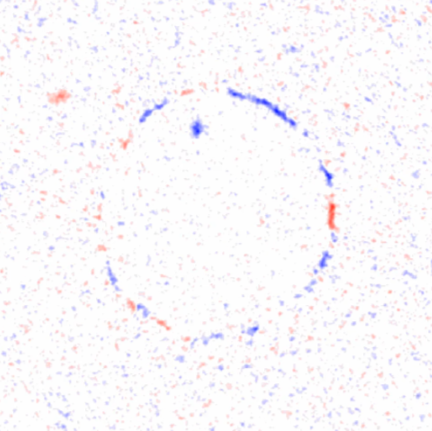}
        \caption{$R^4$}
    \end{subfigure}
    \begin{subfigure}[t]{0.23\textwidth}
        \includegraphics[width=\columnwidth]{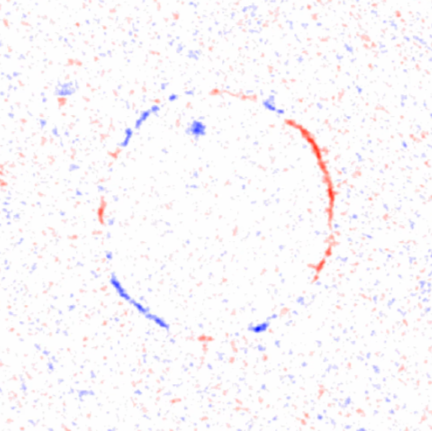}
        \caption{$R^5$}
    \end{subfigure}
    \begin{subfigure}[t]{0.0192\textwidth}
        \includegraphics[height=12\columnwidth, width=\columnwidth]{R1_cb.png}
    \end{subfigure}
    \begin{subfigure}[t]{0.0192\textwidth}
        \includegraphics[height=12\columnwidth]{plus_minus.png}
    \end{subfigure}
    \\[20pt]
    \begin{subfigure}[t]{0.23\textwidth}
        \includegraphics[width=\columnwidth]{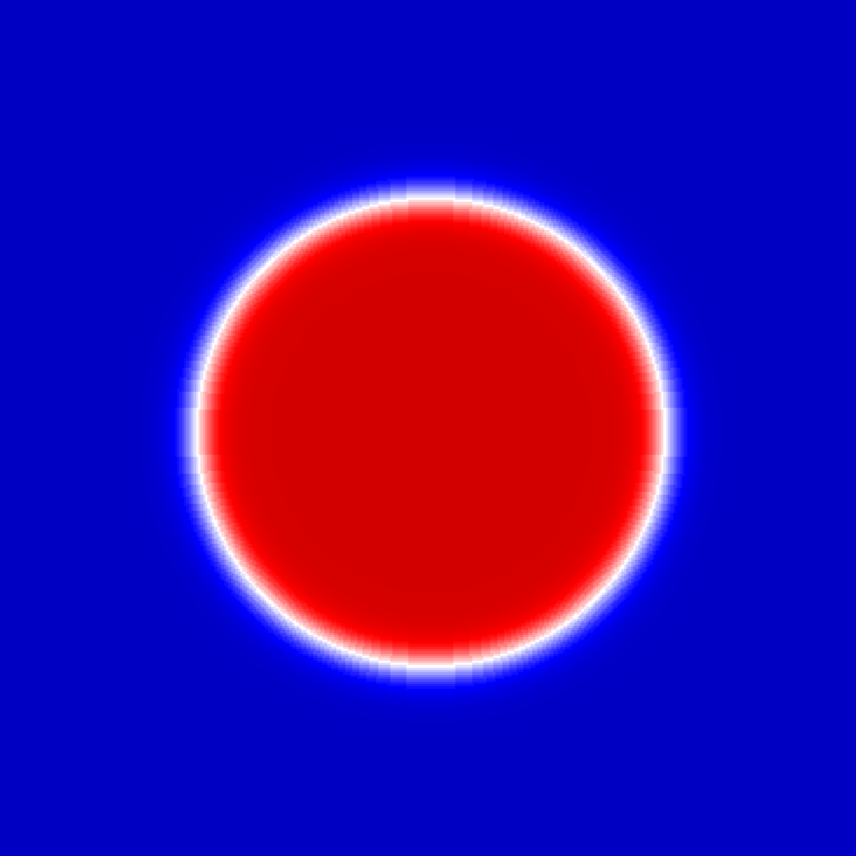}
        \caption{$\phi$}
    \end{subfigure}
    \begin{subfigure}[t]{0.23\textwidth}
        \includegraphics[width=\columnwidth]{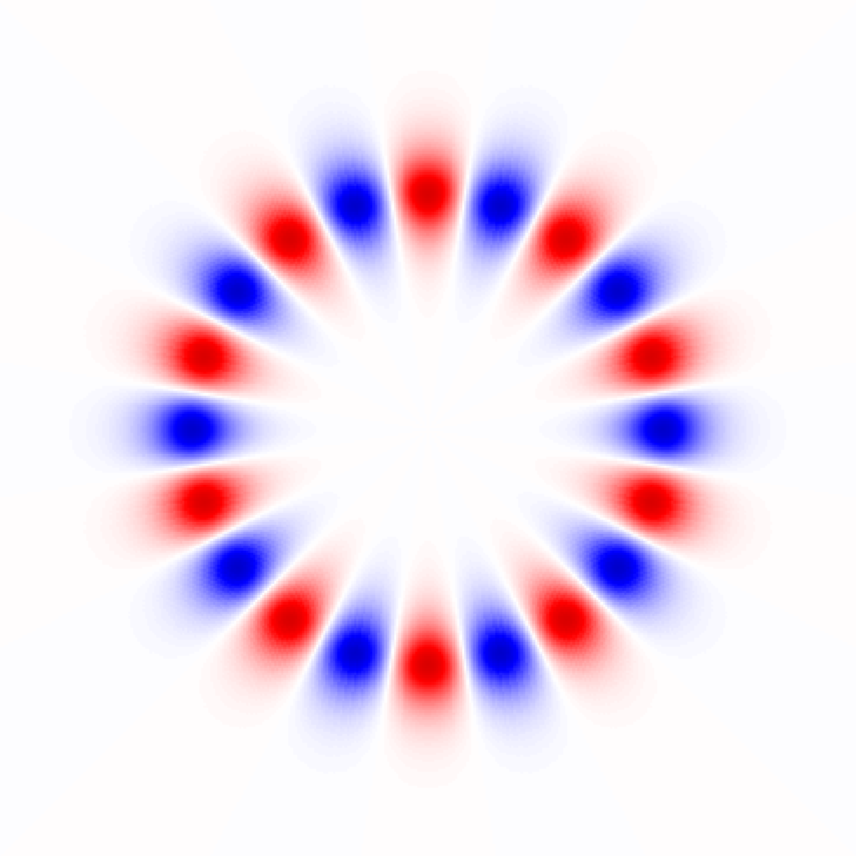}
        \caption{$\text{Re}(\sigma)$}
    \end{subfigure}
    \begin{subfigure}[t]{0.23\textwidth}
        \includegraphics[width=\columnwidth]{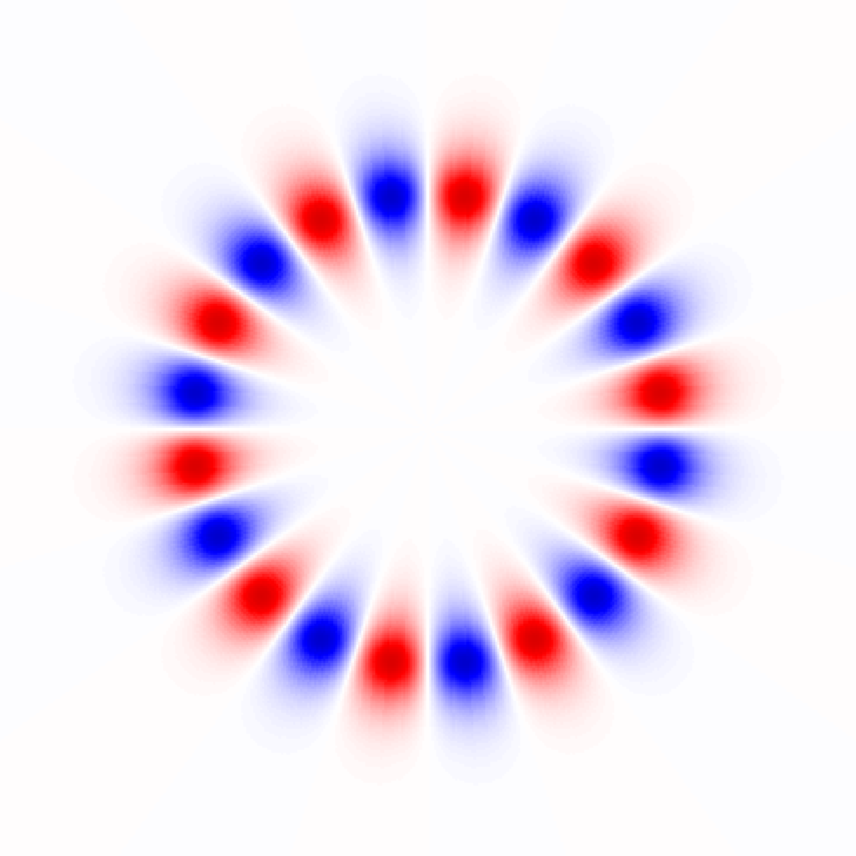}
        \caption{$\text{Im}(\sigma)$}
    \end{subfigure}
    \begin{subfigure}[t]{0.0192\textwidth}
        \includegraphics[height=12\columnwidth, width=\columnwidth]{R1_cb.png}
    \end{subfigure}
    \begin{subfigure}[t]{0.0192\textwidth}
        \includegraphics[height=12\columnwidth]{plus_minus.png}
    \end{subfigure}
    \caption{In the top four panels we present a specific, but typical, realization which created a long-lived loop. In the top leftmost plot we have used a colour scheme similar to Fig.~\ref{fig:states} for the bi-linear combination $R^1$. The next panels along illustrate vector components $R^2, R^4 \text{ and } R^5$ for this realization which can all be seen to wind around the loop. In the bottom two panels we present a circular Kinky Vorton solution with a winding of $N=10$ and radius $R=50$ in the $\mathbb{Z}_2 \times U(1)$ model - it is clear that $\phi$ is very similar to $R^1$ and $\sigma$ to $R^2, R^4 \text{ and } R^5$.}
    \label{fig:KV_2HDM_comp}
\end{figure}

It was already pointed out in ref.~\cite{Battye_2021_Dom_sims} that long-lived loops in simulations up to 1 LCT were reminiscent of Kinky Vortons in the $\mathbb{Z}_2\times U(1)$ model~\cite{Battye_2008_KV}. We have visually examined the configurations in the region $0 < N_{\rm DW} < 500$ at 6 LCTs and indeed there are many characteristics which are similar. In Fig.~\ref{fig:KV_2HDM_comp} we present an approximately circular loop chosen from one of the simulations. It is clear that this is reminiscent of a Kinky Vorton, with the $R^4 \text{ and } R^5$ components being similar to $\text{Re}(\sigma)$ and $\text{Im}(\sigma)$, and $R^2$ exhibiting a further winding condensate structure.

It is known that stable Kinky Vortons exist within the $\mathbb{Z}_2 \times \text{U(1)}$ model. However, the typical amount of charge that localizes upon the domain walls within our RIC simulations was consistently two to three orders of magnitude lower than required for the smallest Kinky Vorton solution found in \cite{Battye_2008_KV}. This could explain why the $\mathbb{Z}_2 \times \text{U(1)}$ model simulations exhibit similar behaviour to those of the simple $\mathbb{Z}_2$ model. We speculate that in the 2HDM simulations the amount of charge and current upon the walls could be closer to that which would be required to form a stable Kinky Vorton. Work is currently in progress to identify these solutions in the 2HDM and to test this hypothesis.

\begin{figure}
    \centering
    \includegraphics[width=0.75\linewidth]{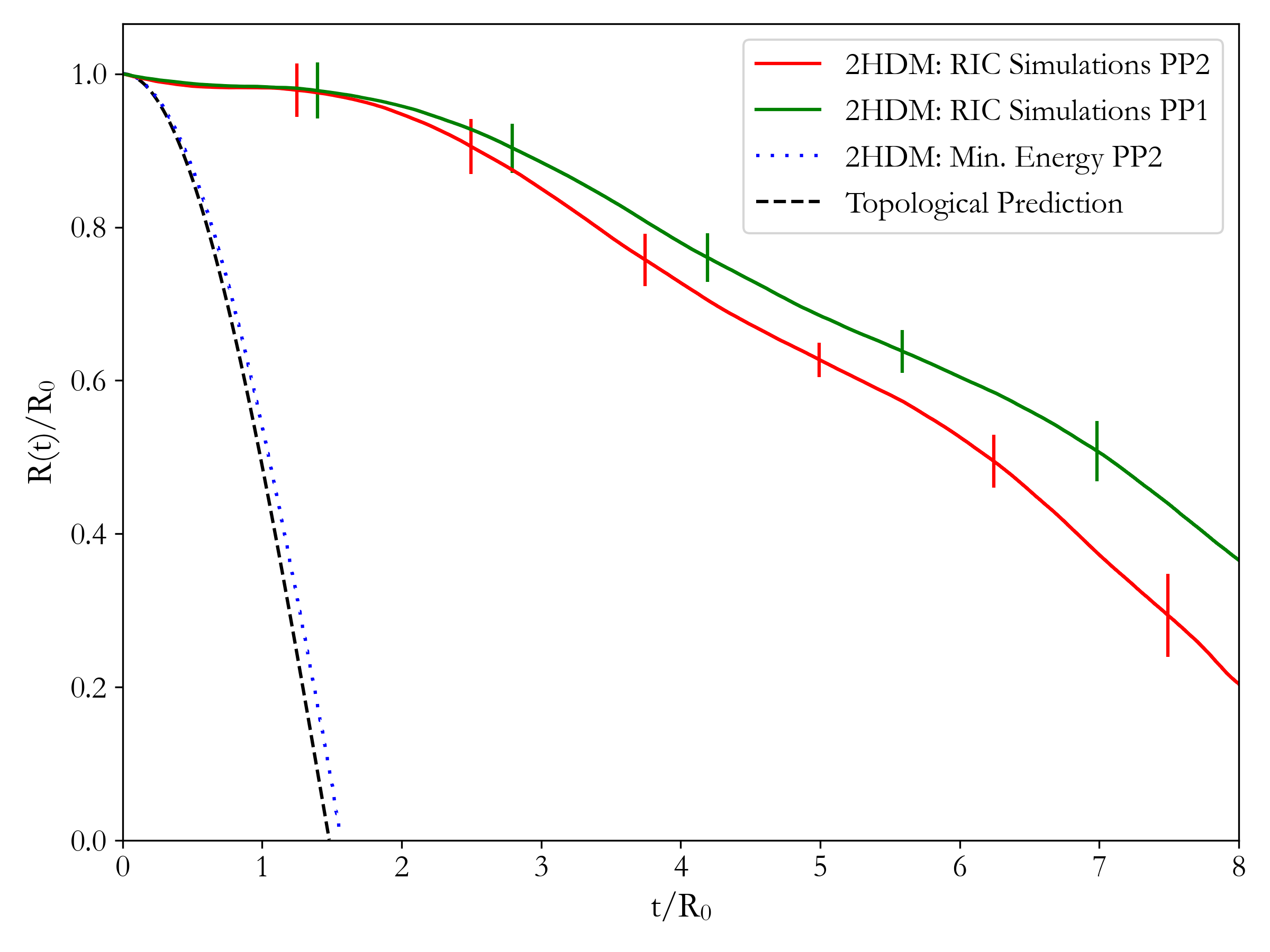}
    \caption{Average decay rates of two different selections of 50 loops of domain wall, PP2 (red line) and PP1 (green line), from 2HDM random simulations compared to the prediction (black dashed-line) of (\ref{eq:predicted_loop_collapse}) and the decay of a constructed loop with no condensate for PP2 (blue dotted-line). Error bars represent one standard deviation of the sample. PP1: $M_H = 800~\text{GeV}, M_A = 500~\text{GeV}, M_{H^\pm} = 400~\text{GeV}$; PP2: $M_H = M_A = M_{H^\pm} = 200~\text{GeV}$}
    \label{fig:rate_decay}
\end{figure}

To further investigate the nature of these long-lived loops within the 2HDM, a selection of realizations were chosen (using their locations within the histogram of Figure \ref{fig:hists_6lct}) such that they contained a long-lived loop. A two dimensional, circular, topological loop collapsing under its own tension has a radius, $R(t)$, which is predicted to decay in time as~\cite{Vilenkin278400}
\begin{equation}
R(t) = R_0 \cos\left(\frac{t}{R_0}\right),
\label{eq:predicted_loop_collapse}
\end{equation}
where $R_0$ is the initial radius. This prediction was confirmed by constructing a circular domain wall of minimum energy configuration, without a condensate, according to the analysis presented in ref.~\cite{Battye_2021_Dom_sims}. All loops taken from realizations with RIC were found to drastically deviate from this prediction, as shown in Fig. \ref{fig:rate_decay}. Such a large deviation clearly indicates a mechanism that slows the collapse, which we suggest here to be due to the existence of a superconducting condensate. Ultimately, all loops collapse meaning that there are no stable Kinky Vortons produced. However, the number of realizations is relatively small and there is significant radiation flowing around in the box which could destabilise the configurations. It is noteworthy that a solution with a stable condensate exists within a different parameter set (PP1) to the one used here (PP2), see ref.~\cite{sassi2023domainwallstwohiggsdoubletmodel} for definitions of PP1 and PP2. However, we found the same behaviour to occur in that regime as the one we have focused on, which is included in Fig. \ref{fig:rate_decay}. Once more, all long-lived loops formed eventually decay, albeit at a marginally slower rate. One thing that is clear from the analysis that we have done is that there are currents flowing along the loops and this can only lead to a reduction in the rate of decay. 

\section{Conclusion}\label{sec:conclusion}

The objective of this letter was to investigate the dynamics of domain walls in the $\mathbb{Z}_2$ symmetric 2HDM in two dimensions. In contrast to previous studies~\cite{Battye_2021_Dom_sims}, which ran up to one LCT in order to measure the scaling properties, we investigated the long term dynamics and in particular tried to use $f_{\rm SD}$ as a diagnostic. As a methodological conclusion we found that method B gave the clearest picture: we were able to deduce that $f_{\rm SD}=0.625\pm 0.003$ for the 2HDM using the parameters chosen here. A by-product of this was that we were able to deduce a similar number for the simple $\mathbb{Z}_2$ and $\mathbb{Z}_2\times U(1)$ models, although the amount of radiation in these models seems to make it more difficult to get stable configurations and hence calculate $f_{\rm SD}$. We have suggested that this is due to only massive radiation being present in these models.

These conclusions are similar to the predictions from the analytic argument presented in ref.~\cite{prob} with any differences being likely related to finite size effects. It was claimed that the analytic prediction applies to any curvature driven coarsening process. Given the different nature of the dynamics discussed here, this has provided an interesting test of this claim. We note that previous numerical work~\cite{PhysRevE.63.036118, Sundaramurthy_2005} has shown that two-dimensional Ising ferromagnets reach a stripe state at a frequency in agreement with the prediction of ref.~\cite{prob}, which demonstrates its applicability to processes with non-relativistic dynamics, and we have now confirmed this for a relativistic case.

This suggests that the 2HDM is no different to any other model in terms of its long term percolation. However, the percolation to the true vacuum seems much slower in the case of the 2HDM which suggests that something more complicated is going on during the relaxation process.  We have presented evidence that the decay of the radius of circular loops is much slower than would be expected if the tension in the wall is the only force acting. We have suggested that this could be due to the existence of superconducting currents on the walls which have the tendency to oppose the tension force. This is supported by visualization of the loops. We did not find any evidence for loops which were entirely stable - that is, Kinky Vortons - and we attribute this, possibly, to the effects of destabilising radiation, but also we cannot be sure, at this stage, if Kinky Vortons are even possible for this set of parameters.

The simulations were done in two spatial dimensions. This is what has allowed us to run so many realizations and hence get such good resolving power on the value of $f_{\rm SD}$. One might ask why such simulations might have any relevance to the case of a three dimensional Universe and to the possible formation of Vortons. There could be many complications in going from two to three spatial dimensions, but for sure they serve as an interesting test case which is numerically tractable. This work motivates further study of (i) domain walls in three spatial dimensions where it would be interesting to see where this phenomena persists, and (ii) also the case of the 2HDM with a $U(1)$ symmetry. Vortices have been studied in models with $U(1)$ symmetry~\cite{Battye_2011_vacTop,Eto:2018tnk,Eto:2021dca,Battye:2024iec}, and recent work has found superconducting vortex solutions~\cite{Battye:2024dvw}. It could be that Vortons~\cite{Davis:1988ij,Lemperiere:2003yt,Battye:2008mm,Battye:2021sji,Battye:2021kbd} or similar configurations can be produced in such a model.

\bibliography{References.bib}
\bibliographystyle{unsrt}

\end{document}